\documentclass[runningheads]{llncs}

\usepackage{floatrow}
\newfloatcommand{capbtabbox}{table}[][\FBwidth]

\usepackage{todonotes}
\usepackage{fixltx2e}
\usepackage{amssymb}
\usepackage{graphicx}
\usepackage{wrapfig}
\usepackage{subfig}
\usepackage{url}
\usepackage{amsmath}
\usepackage{verbatim}
\usepackage{multirow}
\usepackage{paralist}
\usepackage{mathptmx}
\usepackage{times}
\usepackage{dirtree}
\usepackage{tabularx}
\usepackage{tikz}
\usetikzlibrary{arrows,%
                calc,%
                chains,%
                decorations.pathmorphing,%
                decorations.pathreplacing,%
                decorations.shapes,%
                graphs,%
                matrix,%
                positioning,%
                shapes.geometric,%
                shapes,%
                shapes.symbols,%
                trees
   			  }

\usepackage{graphicx}
\usepackage{pgfplots}

\usetikzlibrary{automata,arrows}
\usetikzlibrary{datavisualization}
\usetikzlibrary{datavisualization.formats.functions}
\usetikzlibrary{positioning, calc, backgrounds, arrows}
\usetikzlibrary{shapes.geometric}
\usetikzlibrary{decorations.pathreplacing}
\usetikzlibrary{patterns}

\let\olddefinition\definition
\def\definition{\olddefinition\footnotesize}

\newcommand{\optional}[1]{}

\usepackage[ruled,vlined,boxed,linesnumbered]{algorithm2e}
\SetKwRepeat{Do}{do}{while}
\SetKwInOut{Input}{input}
\SetKwInOut{Output}{output}

\graphicspath{{images/}}

\newcommand{\labels}{\mathcal{A}}

\newcommand{\events}{\mathcal{E}}
\newcommand{\revents}{\mathcal{E_\rho}}
\newcommand{\trace}{t}

\newcommand{\elog}{\mathcal{L}}
\newcommand{\rlog}{\mathcal{L_\rho}}
\newcommand{\dfg}{\mathcal{G}}

\begin{document}

\title{Automated Discovery of Process Models with True Concurrency and  Inclusive Choices}
\author{	Adriano Augusto\inst{1} \and
		    Marlon Dumas\inst{2} \and
			Marcello La Rosa\inst{1}
			 }
\titlerunning{ }
\authorrunning{ }

\institute{
           University of Melbourne, Australia\\
   		\{a.augusto, marcello.larosa\}@unimelb.edu.au
           \and
           University of Tartu, Estonia\\
           marlon.dumas@ut.ee
}

\maketitle \addtocounter{footnote}{-2}

\begin{abstract}
Enterprise information systems allow companies to maintain detailed records of their business process executions. These records can be extracted in the form of event logs, which capture the execution of activities across multiple instances of a business process. Event logs may be used to analyze business processes at a fine level of detail using process mining techniques. Among other things, process mining techniques allow us to discover a process model from an event log -- an operation known as automated process discovery. Despite a rich body of research in the field, existing automated process discovery techniques do not fully capture the concurrency inherent in a business process. Specifically, the bulk of these techniques treat two activities A and B as concurrent if sometimes A completes before B and other times B completes before A. Typically though, activities in a business process are executed in a true concurrency setting, meaning that two or more activity executions overlap temporally. This paper addresses this gap by presenting a refined version of an automated process discovery technique, namely Split Miner, that discovers true concurrency relations from event logs containing start and end timestamps for each activity. The proposed technique is also able to differentiate between exclusive and inclusive choices. We evaluate the proposed technique relative to existing baselines using 11 real-life logs drawn from different industries.
\end{abstract}

\section{Introduction}\label{sec:introduction}

Enterprise information systems, such as Enterprise Resource Planning (ERP) systems, maintain detailed records of each execution of the business processes they support. These records can be extracted in the form of event logs. An event log is a set of event records capturing the execution of activities across a set of instances of a process.

Process mining techniques allow us to exploit event logs in order to analyze business processes at a fine level of detail. Among other things, process mining techniques allow us to discover a process model from an event log -- an operation known as \emph{automated process discovery}. Despite a rich body of research in the field, existing automated process discovery techniques do not fully capture the concurrency inherent in business processes. Indeed, the bulk of automated process discovery techniques operate under an interleaved concurrency model -- a model of concurrency where two events are concurrent if they occur in either order. Specifically, existing techniques treat two activities A and B as concurrent if sometimes A completes before B and other times B completes before A. The interleaved concurrency model is suitable in systems where actions are atomic. However, in a business process, activities have a duration and the execution of two or more activities may overlap temporally. In other words, business processes contain \emph{true concurrency}. The failure of existing automated process discovery techniques to take into account this true concurrency leads them to miss certain concurrency relations. For example, when an activity A always completes before activity B (because B takes longer) even though A and B overlap, existing techniques treat A and B as sequential. If A is then followed by C and C usually completes after B (but overlaps with it), they conclude that A, B and C are sequential, thus missing the observed concurrency. 


This paper addresses this gap by presenting a refined version of an automated process discovery algorithm, namely Split Miner~\cite{augusto2018split}, capable of discovering true concurrency relations from event logs that record both the start and end timestamps of activity executions. The proposed technique, namely Split Miner 2.0, is also able to differentiate between exclusive and inclusive choices. The paper reports on an empirical evaluation that compares Split Miner 2.0 against existing baselines in terms of accuracy and model complexity measures.

The rest of the paper is structured as follows. Section~\ref{sec:background} briefly reviews existing automated process discovery techniques. Section~\ref{sec:approach} introduces the approach to exploit true concurrency for automated process discovery. Section~\ref{sec:evaluation} presents the empirical evaluation while Section~\ref{sec:conclusion} summarizes the findings and further possible extensions.

\section{Background and Related Work}
\label{sec:background}



An \emph{event log} records information about a set of executions of a business processes (a.k.a.\ cases). Concretely, an event log is a chronological sequence of events, each one capturing a state change in the execution of an activity.
As a minimum, each event in a log has three attributes:
the identifier of the process execution (a.k.a. \emph{case ID});
the label (i.e. the process activity the event refers to);
and the timestamp (e.g.\ 10/07/2020 10.43).
Optionally, an event may have other attributes such as the resource who triggered the event, their department, etc.
In this paper, we require that at least one fourth attribute is attached to each event,
namely the \emph{life-cycle transition}. For a given event, this attribute indicates what state-change the referenced activity has undergone.
The life-cycle of an activity captures all the states in an activity execution and their possible transitions.
In general, one could observe very complex life-cycles, including states such as created, assigned, started, suspended, etc.
In this paper, we adopt a simple life-cycle model wherein an activity execution can be in one of two states: \emph{start} (i.e. the activity execution started); and \emph{end} (i.e the activity execution ended).



\begin{figure}
\begin{floatrow}
\ffigbox{%
\centering
  \includegraphics[width=0.5\textwidth]{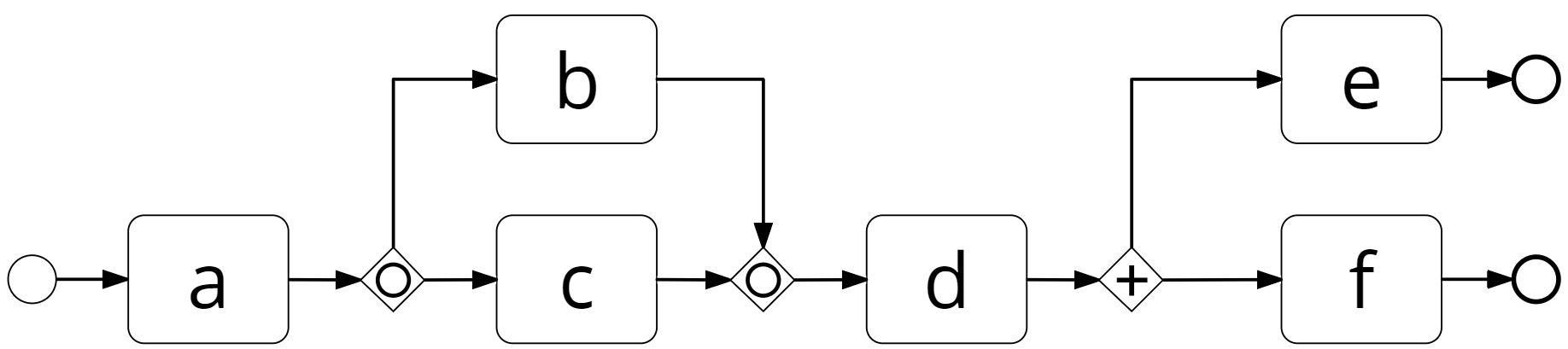}
}{%
  \caption{Process model example.}\label{fig:modelex}
}
\capbtabbox{%
\centering
	{\scriptsize{
    \begin{tabular}{c|c|c|c}
    
    \textbf{Case-ID}
    & \textbf{Activity}
    & \textbf{Life-cycle}
    & \textbf{Timestamp}\\\hline
    
1	&	a	& start	& 2020-07-08 10.03\\
2	&	a	& start	& 2020-07-08 10.42\\	
1	&	a	& end	& 2020-07-08 10.57\\	
2	&	a	& end	& 2020-07-08 11.21\\	
1	&	b	& start	& 2020-07-08 13.29\\	
1	&	c	& start	& 2020-07-08 14.13\\	
2	&	b	& start	& 2020-07-08 15.22\\	
2	&	b	& end	& 2020-07-09 10.24\\	
1	&	b	& end	& 2020-07-09 10.37\\	
2	&	d	& start	& 2020-07-09 11.13\\
2	&	d	& end	& 2020-07-09 12.28\\	
1	&	c	& end	& 2020-07-09 12.53\\\hline

	\end{tabular}
	}}
}{%
  \caption{Event log example.}\label{tab:logex}
}
\end{floatrow}
\vspace{-1.0\baselineskip}
\end{figure}

Event logs can be exploited for different types of analysis including conformance checking, process performance mining, and automated process discovery\cite{ProcessMiningBook2}.
In this paper, we focus on the latter. The goal of automated process discovery is to discover a process model (such as the one in Figure~\ref{fig:modelex}) by analysing an event log such as the one in Table~\ref{tab:logex} (the latter is just an extract and not a full log).

The quality of an automatically discovered process model is traditionally assessed over four dimensions:
fitness -- the amount of process behaviour recorded in the event log that can be replayed by the process model;
precision -- the amount of behaviour captured by the process model that can be found in the event log;
generalization -- the amount of behaviour captured by the process model that even not being observed in the event log is likely to belong to the original process;
and simplicity -- quantifying how difficult is to understand the process model.
Furthermore, a process model should be sound.
The notion of \emph{soundness} has been defined on Workflow nets~\cite{AalstHHSVVW11} as a correctness criterion, and is also applicable to BPMN models. Formulated on BPMN models, soundness encompasses three properties:
i) every process instance eventually reaches the end event (no deadlocks);
ii) no end event is reached more than once during a process execution (proper completion);
iii) each process activity is triggered in at least one process execution (no dead activities).

A recent literature review of automated process discovery algorithms~\cite{augusto2018automated}
showed that only few algorithms stand out for accuracy and performance among those outputting procedural process models.
Specifically, Inductive Miner (IM)~\cite{leemans2014infrequent}, Evolutionary Tree Miner (ETM)~\cite{BuijsDV12}, and Split Miner (SM)~\cite{augusto2018split}.
IM and ETM are known to discover process models that are either highly fitting or precise, discovering simple, block-structured and sound process models,
while SM focuses on maximizing both fitness and precision at the cost of simplicity, structuredness, and in rare cases compromising the soundness of the process models~\cite{augusto2018automated,augusto2018split}.
However, of these three automated process discovery algorithms, only IM provides a variant that takes into account the activities' life-cycle when discovering a process model. IM life-cycle variant~\cite{leemans2016using} analyses the activities' life-cycles to distinguish between concurrency and interleaving relations.

Past studies that investigated the problem of discovering control-flow relations between activities by leveraging life-cycle information or execution times include:
(1) a simple algorithm~\cite{schimm2004mining} for discovering block-structured process models from complete and noise-free event logs;
(2) an extension of the $\alpha$-algorithm, i.e. the $\beta$ algorithm~\cite{wen2009novel}; 
(3) an extension of Heuristics Miner~\cite{burattin2010heuristics};
and (4) the work of Senderovich et al. which explores process performance modelling via temporal network representation~\cite{senderovich2017temporal}. The first one is limited to noise-free log. The second and third are based on underlying algorithms that produce unsound and inaccurate models when applied to real-life event logs, as shown in~\cite{augusto2018automated}. The fourth approach is not geared to discovering process models but rather targets the problem of performance mining.



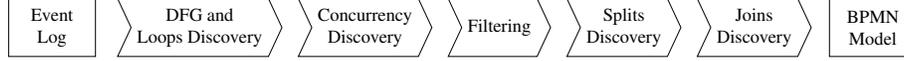
\begin{figure}[htb]
 \centering\footnotesize
 \resizebox{1\textwidth}{!}{
 \begin{tikzpicture}
   \tikzset{
     bound/.style={
       draw,
       minimum height=1cm,
       minimum width=1.5cm,
       align=center
     },
     arrow/.style={
       draw,
       minimum height=1cm,
       shape=signal,
       signal from=west,
       signal to=east,
       signal pointer angle=130,
       align=center
     }
   }
   \node[bound] (a) { Event\\  Log};
   \begin{scope}[start chain=transition going right, node distance=0.5cm]
     \node[arrow,right=0.6cm of a, on chain] (b) { DFG and\\  Loops Discovery};
     \node[arrow,right=0.3cm of b, on chain] (c) { Concurrency\\  Discovery};
     \node[arrow,right=0.3cm of c, on chain] (d) { Filtering};
     \node[arrow,right=0.3cm of d, on chain] (e) { Splits\\  Discovery};
     \node[arrow,right=0.3cm of e, on chain] (f) { Joins\\  Discovery};
   \end{scope}
   \node[bound,right=0.3cm of f] { BPMN\\  Model};
 \end{tikzpicture}
 }
 \caption{Overview of the Split Miner approach~\cite{augusto2018split}.}\label{fig:splitminer}
 \end{figure}

In this paper, we extend the SM algorithm, which has been shown to produce accurate and (generally) sound process models over real-life logs.
Figure~\ref{fig:splitminer} shows an overview of how SM discovers a process model from an event log.
Given an input event log, SM operates over five steps:
i) discover the directly-follows graph (DFG) and loops from the event log;
ii) analyse the DFG for discovering concurrency relations;
iii) filter the DFG by removing the infrequent behaviour;
iv) discover the split gateways;
v) discover the join gateways. 
Each step is a standalone operation based on tailored algorithms~\cite{augusto2018split},
such a modular approach allows the replacement of any step with alternative methods.
In this paper, we show how we updated the first, second, and fifth steps 
to discover true concurrency and inclusive choices, and reduce the chances of producing unsound process models via heuristics.
\vspace{-2.0\baselineskip}

\section{Approach}
\label{sec:approach}

In this section, we describe how we redesigned the first two steps of the Split Miner original approach~\cite{augusto2018split}
and integrated in the last step two heuristics to repair models that are unsound due to improper completion and identify inclusive relations between activities, enabling the discovery of OR-splits.

\vspace{-0.5\baselineskip}
\subsection{Refined Directly-follows Graph Discovery}

Given an event log, the first step performed by Split Miner is to sequentially read the events and build the directly-follows graph (DFG). Although this operation is straightforward, its output strictly depends on how the event log and the DFG are defined.
Definitions~\ref{def:elog},~\ref{def:dfr0}, and~\ref{def:dfg} capture the notion of DFG used in the original Split Miner.

\begin{definition}[Event Log as in~\cite{augusto2018split}]\label{def:elog}
Given a set of process activity labels $\labels$,
an \emph{event log} $\elog$ is a multiset of traces, where a trace $t \in \elog$ 
is a sequence of activity labels $\trace = \langle a_1, a_2, \dots, a_k \rangle$, with $a_i \in \labels, 1 \leq i \leq k$. 
In addition, we use the notation $a \in \labels$ to refer an activity $a$ that belongs to a generic trace $t \in \elog$.
\footnote{For simplicity, we use the term activity to refer to its label.}
\end{definition}
\begin{definition}[Directly-Follows Relation as in~\cite{augusto2018split}]\label{def:dfr0}
Given an event log $\elog$ and two process activities $a_x, a_y \in \labels$,
we say that activity $a_y$ directly-follows activity $a_x$, with notation $a_x \leadsto a_y$,
if and only if (iff) $\exists~ \langle a_1, a_2, \dots, a_k \rangle \in~\elog \mid a_i = a_x \wedge a_j = a_y \wedge j = i+1 \wedge 0 < i < n$.
\end{definition}
\begin{definition}[Directly-Follows Graph as in~\cite{augusto2018split}]\label{def:dfg}
Given an event log $\elog$, its \emph{Directly-Follows Graph (DFG)} is a directed graph $\dfg = (N, E)$, where $N$ is the non-empty set of nodes, where each node represents a unique activity $a \in \elog$ and there exists a bijective function $\lambda : N \mapsto \labels$ such that $\lambda(n)$ retrieves the activity $n$ refers to;
and $E$ is the set of edges capturing the directly-follows relations of the activities observed in $\elog$, $E = \{(n, m) \in N \times N \mid \lambda(n) \leadsto \lambda(m)\}$.
\end{definition}
To capture the activities' lifecycle information, we refine the concept of event log.
\begin{definition}[Refined Event Log]\label{def:relog}
Given a set of events $\events$,
a \emph{refined event log} $\rlog$ is a multiset of traces, where a trace $t \in \rlog$ 
is a sequence of events $\trace = \langle e_1, e_2, \dots, e_k \rangle$,
with $e_i \in \events, 1 \leq i \leq k$. 
Each event $e \in \revents$ is a tuple $e=(l, p, t)$, where
$l \in \labels$ is the process activity the event refers to, retrieved with the notation $e^l$;
$p \in \{\mathit{start},\mathit{end}\}$ is the state of the life-cycle of activity $l$, retrieved with the notation $e^p$;
and $t$ is the timestamp of the event, retrieved with the notation $e^t$.
\end{definition}
While redefining the event log to capture the activities' life-cycle information is intuitive and follows from its original definition~\cite{ProcessMiningBook2},
the same does not apply for the DFG. Indeed, more than one approach could be used to generate a DFG from a refined event log.
The simplest approach would be to disregard all the events of a specific state of an activity life-cycle,
for example, we could remove from $\rlog$ all the events $e \in \rlog \mid e^p = \mathit{start}$ or all the events $e \in \rlog \mid e^p = \mathit{end}$. Then, the refined event log would turn into an event log (Definition~\ref{def:elog}) and the DFG would be constructed according to Definition~\ref{def:dfg}, but this would be equivalent to discarding the activities' lifecycle information.

An alternative approach was proposed by Leemans et al.~\cite{leemans2016using} and incorporated into a variant of the Inductive Miner that takes into account lifecycle transitions, herein called Inductive Miner Lifecycle (IM-lc).  
According to~\cite{leemans2016using}, an activity $a_y$ directly-follows an activity $a_x$ if any of the life-cycle states of activity $a_y$ is observed after any of the life-cycle states of activity $a_x$ in the same trace and between the two observations no activity completes the execution of its full life-cycle (see Definition~\ref{def:dfr1}).
\begin{definition}[Directly-Follows Relation as in~\cite{leemans2016using}]\label{def:dfr1}
Given a refined event log $\rlog$ and two process activities $a_x, a_y \in \labels$, the relation $a_x \leadsto a_y$ holds iff
$\exists~ \langle e_1, e_2, \dots, e_k \rangle \in~\rlog \mid {e^l}_i = a_x \wedge {e^l}_j = a_y \wedge i < j \wedge \nexists n,m \in~]~i,j~[~\mid~n < m ~\wedge {e^p}_n = \mathit{start} \wedge {e^p}_m = \mathit{end} \wedge {e^l}_n = {e^l}_m$. 
\end{definition}
According to Definition~\ref{def:dfr1}, a directly-follows relation would hold between two activities whose life-cycles overlap (i.e. the start-state of an activity is observed between the start-state and the end-state of another activity).  
While this is important and useful for IM-lc to discover concurrency relations~\cite{leemans2014infrequent},
it would not be beneficial for Split Miner, since Split Miner requires to remove the directly-follows relations between activities that are considered concurrent~\cite{augusto2018split}. Consequently, we are interested in discarding directly-follows relations of activities whose life-cycles overlap. We redefine the directly-follows relation of activities observed in a refined event log as follows. An activity $a_y$ directly-follows an activity $a_x$ if the start-state of the life-cycle of activity $a_y$ is observed after the end-state of the life-cycle of activity $a_x$ and no end-state of other activities are observed in between (see Definition~\ref{def:dfr2}).
\begin{definition}[Directly-Follows Relation]\label{def:dfr2}
Given a refined event log $\rlog$ and two process activities $a_x, a_y \in \labels$,
the relation $a_x \leadsto_r a_y$ holds iff
$\exists~ \langle e_1, e_2, \dots, e_k \rangle \in~\rlog \mid {e^l}_i = a_x \wedge {e^l}_j = a_y \wedge {e^p}_i = \mathit{end} \wedge {e^l}_j = \mathit{start} \wedge 1 \leq i < j \leq k \wedge \nexists n \in~]~i,j~[~\mid {e^p}_n = \mathit{end}$.
\end{definition}
\vspace{-1.5\baselineskip}
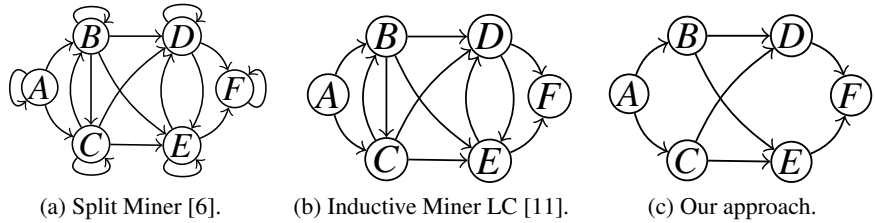
\begin{figure}[h]
\vspace{-1.5\baselineskip}
\tikzset{loop above/.style={min distance=2mm,in=45,out=135,looseness=3}}
\tikzset{loop below/.style={min distance=2mm,in=-45,out=-135,looseness=3}}
\tikzset{loop right/.style={min distance=2mm,in=45,out=-45,looseness=3}}
\tikzset{loop left/.style={min distance=2mm,in=-135,out=135,looseness=3}}
\centering
\resizebox{0.99\textwidth}{!}{
\subfloat[Split Miner~\cite{augusto2018split}.\label{fig:dfg0}]{
\centering
\resizebox{0.32\textwidth}{!}{
 \tikzstyle{round} = [draw, circle, scale=1.0, inner sep=0.4] 
\begin{tikzpicture}[node distance=-0.3pt]
  \node[round] (a) {$A$}; 
  \node[round, above right=1em of a] (b) {$B$}; 
  \node[round, below=2em of b] (c) {$C$}; 
  \node[round, right=1.5em of b] (d) {$D$}; 
  \node[round, below=2em of d] (e) {$E$}; 
  \node[round, below right=1em of d] (f) {$F$};
  
  \draw[->] (a) edge[loop left] node[auto] {} (a);
  \draw[->] (a) edge[bend left=25] node[auto] {} (b);
  \draw[->] (a) edge[bend right=25] node[auto] {} (c);
  
  \draw[->] (b) edge[loop above] node[auto] {} (b);
  \draw[->] (b) edge node[auto] {} (c);
  \draw[->] (b) edge node[auto] {} (d);
  \draw[->] (b) edge[bend right=15] node[auto] {} (e);
  
  \draw[->] (c) edge[loop below] node[auto] {} (c);
  \draw[->] (c) edge[bend left=30] node[auto] {} (b);
  \draw[->] (c) edge[bend left=15] node[auto] {} (d);
  \draw[->] (c) edge node[auto] {} (e);
  
  \draw[->] (d) edge[loop above] node[auto] {} (d);
  \draw[->] (d) edge[bend left=30] node[auto] {} (e);
  \draw[->] (d) edge[bend left=25] node[auto] {} (f);
  
  \draw[->] (e) edge[bend left=30] node[auto] {} (d);
  \draw[->] (e) edge[loop below] node[auto] {} (e);
  \draw[->] (e) edge[bend right=25] node[auto] {} (f);
  
  \draw[->] (f) edge[loop right] node[auto] {} (f);
\end{tikzpicture}
  }}
  
\hfill

\subfloat[Inductive Miner LC~\cite{leemans2016using}.\label{fig:dfg1}]{
\resizebox{0.32\textwidth}{!}{
 \tikzstyle{round} = [draw, circle, scale=1.0, inner sep=0.4] 
\begin{tikzpicture}[node distance=-0.3pt]
  \node[round] (a) {$A$}; 
  \node[round, above right=1em of a] (b) {$B$}; 
  \node[round, below=2em of b] (c) {$C$}; 
  \node[round, right=1.5em of b] (d) {$D$}; 
  \node[round, below=2em of d] (e) {$E$}; 
  \node[round, below right=1em of d] (f) {$F$};
  
  \draw[->] (a) edge[bend left=25] node[auto] {} (b);
  \draw[->] (a) edge[bend right=25] node[auto] {} (c);
  
  \draw[->] (b) edge node[auto] {} (c);
  \draw[->] (b) edge node[auto] {} (d);
  \draw[->] (b) edge[bend right=15] node[auto] {} (e);
  
  \draw[->] (c) edge[bend left=30] node[auto] {} (b);
  \draw[->] (c) edge[bend left=15] node[auto] {} (d);
  \draw[->] (c) edge node[auto] {} (e);
  
  \draw[->] (d) edge[bend left=30] node[auto] {} (e);
  \draw[->] (d) edge[bend left=25] node[auto] {} (f);
  
  \draw[->] (e) edge[bend left=30] node[auto] {} (d);
  \draw[->] (e) edge[bend right=25] node[auto] {} (f);
  
\end{tikzpicture}
  }}
  
\hfill
  
\subfloat[Our approach.\label{fig:dfg2}]{
\resizebox{0.32\textwidth}{!}{
 \tikzstyle{round} = [draw, circle, scale=0.9, inner sep=0.4] 
\begin{tikzpicture}[node distance=-0.3pt]
  \node[round] (a) {$A$}; 
  \node[round, above right=1em of a] (b) {$B$}; 
  \node[round, below=2em of b] (c) {$C$}; 
  \node[round, right=1.5em of b] (d) {$D$}; 
  \node[round, below=2em of d] (e) {$E$}; 
  \node[round, below right=1em of d] (f) {$F$};
  
  \draw[->] (a) edge[bend left=25] node[auto] {} (b);
  \draw[->] (a) edge[bend right=25] node[auto] {} (c);
  
  \draw[->] (b) edge node[auto] {} (d);
  \draw[->] (b) edge[bend right=15] node[auto] {} (e);
  
  \draw[->] (c) edge[bend left=15] node[auto] {} (d);
  \draw[->] (c) edge node[auto] {} (e);
  
  \draw[->] (d) edge[bend left=25] node[auto] {} (f);
  
  \draw[->] (e) edge[bend right=25] node[auto] {} (f);
  
\end{tikzpicture}
  }}   

}  
\caption{Examples of discovered DFGs by applying Definition~\ref{def:dfr0}, \ref{def:dfr1}, and \ref{def:dfr2} (left to right).}\label{fig:dfgs}
\end{figure} 
\vspace{-1.5\baselineskip}

The new version of Split Miner we propose in this paper relies on Definition~\ref{def:dfr2}.
Depending on the definition of directly-follows relation that one adopts when generating the DFG, one may discover very different DFGs.
As an example, let us consider the following refined event log (captured as a collection of traces, where each event is represented as the activity it refers to -- including its life-cycle state as subscript, $s$ standing for \emph{start} and $e$ standing for \emph{end}):
$\rlog_x=$\\
$\{\langle A_s, A_e, B_s, C_s, C_e, B_e, E_s, D_s, D_e, E_e, F_s, F_e\rangle,
\langle A_s, A_e, B_s, C_s, B_e, C_e,  E_s, D_s, E_e, D_e,  F_s, F_e\rangle,$
$\langle A_s, A_e, C_s, B_s, B_e, C_e,  D_s, E_s, D_e, E_e, F_s, F_e\rangle,
\langle A_s, A_e, C_s, B_s, C_e, B_e,  D_s, E_s, E_e, D_e, F_s, F_e\rangle\}$; 
Figure~\ref{fig:dfgs} shows the DFGs discovered from the $\rlog_x$ by applying
Definition~\ref{def:dfr0} (original Split Miner approach),
Definition~\ref{def:dfr1} (Inductive Miner life-cycle approach),
and Definition~\ref{def:dfr2} (this paper approach).
\vspace{-0.5\baselineskip}
\subsection{Refined Concurrency Discovery}

The second step of the original Split Miner that we redesigned is the concurrency discovery.
Split Miner relies on a simple heuristic to discover concurrency, precisely,
given a DFG and two activities $A, B \in \labels$ such that neither $A$ nor $B$ is a self-loop, 
$A$ and $B$ are assumed concurrent iff three conditions are true:
$A$ directly-follows $B$ and $B$ directly-follows $A$ (Relation~\ref{eq2});
$A$ and $B$ do not form a short-loop (Relations~\ref{eq3} and~\ref{eq31});
the frequency of the two directly-follows relations $A\leadsto B$ and $B\leadsto A$ is similar (Relation~\ref{eq3}).\footnote{The frequency of a directly-follows relation is the number of times the relation is observed.}

\vspace{-1.0\baselineskip}
\begin{equation}
A\leadsto B \wedge B\leadsto A \label{eq2}
\end{equation}
\begin{equation}
\nexists \langle a_1, a_2, \dots, a_k \rangle \in \elog \mid a_i = A \wedge a_{i+1} = B \wedge a_{i+2} = A~\wedge~i\in[1,k-2]\label{eq3}
\end{equation}
\begin{equation}
\nexists \langle a_1, a_2, \dots, a_k \rangle \in \elog \mid a_i = B \wedge a_{i+1} = A \wedge a_{i+2} = B~\wedge~i\in[1,k-2]\label{eq31}
\end{equation}
\begin{equation}
\frac{\left| \left| A\leadsto B \right| - \left| B\leadsto A \right| \right|}{\left| A\leadsto B \right| + \left| B\leadsto A \right|} < \varepsilon \quad (\varepsilon \in [0,1]) \label{eq4}
\end{equation}

The simplicity of the concurrency oracle of Split Miner derives from the simplicity of the input event log (see Definition~\ref{def:elog}).
However, when receiving as input a refined event log (Definition~\ref{def:relog}), it is possible to identify true concurrency 
by focusing on activities whose life-cycles overlap and are hence truly executed concurrently (e.g. by different process resources). 
Consequently, we redefine the concurrency discovery oracle as follows.
Given two activities $A, B \in \labels$ and a refined event log $\rlog$,
we say $A$ and $B$ are concurrent if the following relation holds:

\begin{equation}
2 \cdot \frac{\left| A \asymp B \right|}{\left| A \right| + \left| B \right|} \geq \varepsilon \quad (\varepsilon \in [0,1]) \label{eq41}
\end{equation}

where
$\left| A \asymp B \right|$ is the total number of observations of overlapping life-cycles of $A$ and $B$ in $\rlog$;
$\left| A \right|$ and $\left| B \right|$ are respectively the total number of complete life-cycle\footnote{E.g. including start and end states.} observations of activity $A$ and activity $B$ in $\rlog$;
and $\varepsilon$ is an arbitrary variable (given as input parameter) defining the minimum percentage of times that the two activities' life-cycles are required to overlap to assume the two activities concurrent.
In particular, when  $\varepsilon = 1$ our notion of concurrency is equivalent to the notion of \emph{strong simultaneousness} defined by Van der Werf et al.~\cite{van2013mining} as well as Allen's interval relations~\cite{allen1983maintaining} of \emph{overlaps}, \emph{contains}, \emph{starts}, and \emph{is finished by}.
While for any other value of $\varepsilon > 0$ it is equivalent to a parametrized notion of \emph{weak simultaneousness}~\cite{van2013mining}.
Given that real-life event logs often contain noise and infrequent process behaviour, requiring $\varepsilon = 1$ would be very restrictive and may lead to the discovery of no concurrent activities. 

Although both our approach and IM-lc infer concurrency relations between activities from the observation of overlapping life-cycles, we rely on an heuristic before validating the concurrency relations (i.e. Equation~\ref{eq41}) -- in-line with the original Split Miner; while IM-lc assumes the information contained in the log to be valid a priori (this is mitigated by another extension of IM-lc that embeds a filtering technique~\cite{leemans2016using}).

\vspace{-0.5\baselineskip}
\subsection{Heuristic Improvement}

\begin{figure}[htb]
\vspace{-2.5\baselineskip}
	\centering
	\subfloat[Model with improper completion.]{
		\includegraphics[width=0.42\textwidth]{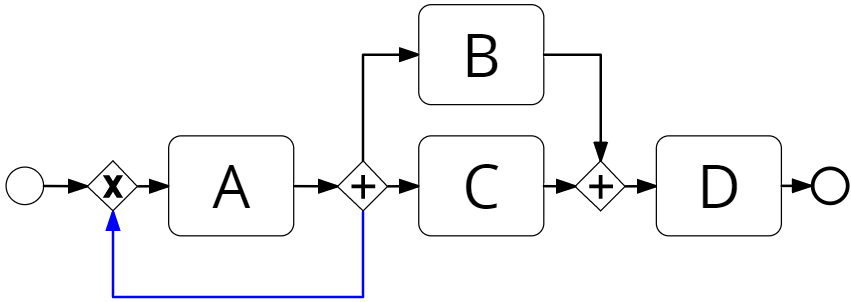}\label{fig:heur0}
	}
	\hfill
	\subfloat[Model after applying heuristic.]{
		\includegraphics[width=0.48\textwidth]{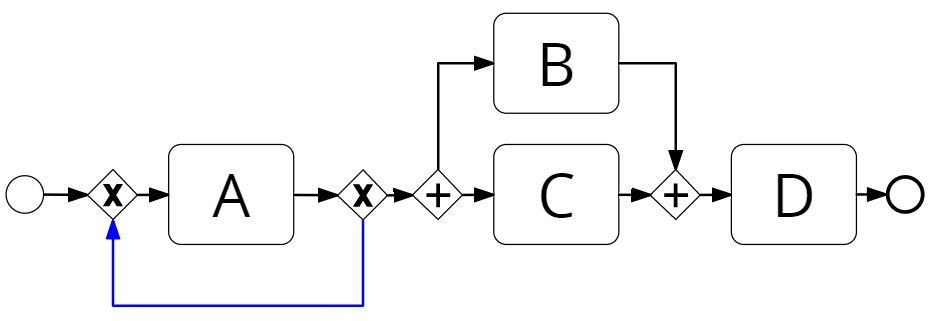}\label{fig:heur1}
	}
\caption{Heuristic removal of improper completion generated by loops.}
\label{fig:loopheur}
\end{figure}

Although Split Miner guarantees to discover sound acyclic process models and \emph{deadlock-free} cyclic process models with \emph{no dead activities},
for cyclic process models it does not guarantee \emph{proper completion}. However, it is possible to reduce the chances to discover process models exhibiting improper completion by applying the following heuristic: for each AND-split gateway in a process model with an outgoing edge that is a loop-edge (leading to a topologically deeper node of the process model), we create a preceding XOR-split gateway and set this latter as source of the loop-edge.
Figure~\ref{fig:loopheur} intuitively show how the heuristic operates, the loop-edge is highlighted in blue and, in general, activities could be present in the loop-edge.

\vspace{-1.5\baselineskip}

\begin{figure}[h]
	\centering
	\subfloat[Before.]{
		\includegraphics[width=0.4\textwidth]{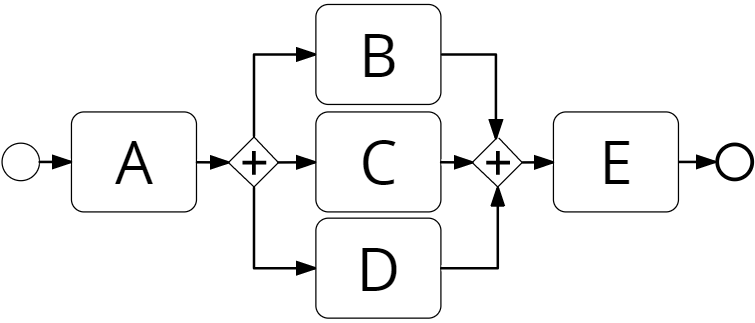}\label{fig:heur2}
	}
	\hfill
	\subfloat[After.]{
		\includegraphics[width=0.4\textwidth]{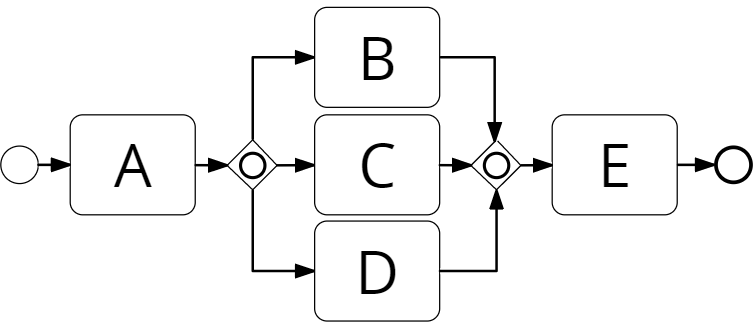}\label{fig:heur3}
	}
\caption{Heuristic identification of OR-split gateways.}
\label{fig:andheur}
\end{figure}

Lastly, we integrated an heuristic to discern between concurrency and inclusive relations, in other words identifying when an AND-split gateway is a candidate OR-split gateway. This second heuristic operates as follows. For each AND-split gateway in a process model, we consider all the successor activities and we check pairwise whether there exist traces where the pair of activities are mutually exclusive (i.e. one of the two activities is executed but not the other). Then, if the majority of the pairs of activities are both mutually exclusive and concurrent in different traces,\footnote{With at least one observation of mutual exclusiveness every two observations of concurrency or vice-versa.} we turn the AND-split gateway into an OR-split gateway and we update accordingly the OR-join gateway. 
As an example, let us consider the model in Figure~\ref{fig:heur2} and the event log
$\rlog_y= \{\langle A_s, A_e, B_s, C_s, D_s, B_e, D_e, C_e, E_s, E_e\rangle^3,
\langle A_s, A_e, C_s, D_s, C_e, D_e, E_s, E_e\rangle^2,$\\
$\langle A_s, A_e, B_s, D_s, D_e, B_e, E_s, E_e\rangle,\}$; 
$B$ and $C$ are observed three times concurrently and three times are mutually exclusive,
$B$ and $D$ are observed four times concurrently and two times mutually exclusive,
$C$ and $D$ are observed five times concurrently and one mutually exclusive.
Given that two pairs of activities out of three ($B,D$ and $B,C$) are eligible for inclusiveness, we turn the AND gateways into OR gateways (Figure~\ref{fig:heur3}).

\vspace{-0.5\baselineskip}
\section{Evaluation}\label{sec:evaluation}

In this section, we present an empirical evaluation that compares Split Miner 2.0 (SM\textsubscript{2.0}) with three state-of-the-art automated process discovery algorithms: the original Split Miner~\cite{augusto2018split} (SM), the Inductive Miner Lifecycle (IM-lc)~\cite{leemans2014infrequent} including its infrequent behaviour filter~\cite{leemans2016using},
and the most recent version of IM, namely IMfa~\cite{leemans2020information}.  

\vspace{-0.5\baselineskip}
\subsection{Dataset and Setup}

As testing dataset, we selected eleven real-life event logs (L1-L11) containing activity lifecycle information. The logs were sourced from companies operating in different fields (e.g.\ insurance, manufacturing, banking) and geographic areas (i.e.\ Europe and Australia).
Given that these logs are not publicly available, we added a publicly available simulated event log known as the ``Repair example'' (R-Log),\footnote{\url{http://www.promtools.org/prom6/downloads/example-logs.zip}} which also contains activity lifecycle information. We did not include the BPIC12 and BPIC17 logs simply because the former does not have any overlapping lifecycle, and for the latter both SM and SM\textsubscript{2.0} produced the same model, which was analysed in previous studies~\cite{augusto2019measuringapproach,augusto2018split,augusto2018automated}.

Table~\ref{tab:logs} displays the characteristics of the event logs, highlighting their variety,
with logs containing short to long traces (length 2 to 1,230),
a wide range of distinct traces (0.28\% to 96.55\%) and distinct events (6 to 80),
as well as notable differences in the total number of traces (37 to 70,512) and events (1,156 to 830,522).
The lifecycle information for each activity recorded in these event logs was complete,
i.e.\ the start and end events were recorded for each activity.

From each log, we discovered a process model with SM\textsubscript{2.0}, SM, and IM-lc,
and compared the quality of the discovered models over three quality measures: fitness, precision, and simplicity.
Several methods have been proposed for measuring fitness and precision of an automatically discovered process model~\cite{syring2019evaluating}. In this paper we use two methods, the one proposed by Adriansyah et al.~\cite{adriansyah2012alignment,adriansyah2015measuring} (alignment-based accuracy) and the one proposed by Augusto et al.~\cite{augusto2019measuringapproach} (Markovian accuracy).
As proxy for simplicity we use the following three metrics~\cite{mendling2008metrics}:
\emph{Size} -- the total number of nodes of a process model;
\emph{Control-flow complexity} (CFC) -- the amount of branching induced by the split gateways in the process model;
\emph{Structuredness} -- the percentage of nodes located inside a single-entry single-exit fragment of the process model.

\begin{table}
\vspace{-1.0\baselineskip}
	\centering
	{\scriptsize{
\bgroup
\setlength{\tabcolsep}{8pt}
    \begin{tabular}{c|cc|cc|ccc}
    
    \textbf{Event}
    & \textbf{Total}
    & \textbf{Distinct}
    & \textbf{Total}
    & \textbf{Distinct}
    & \multicolumn{3}{c}{\textbf{Trace Length}}
    \\\cline{6-8}
    \textbf{Log}
    & \textbf{Traces}
    & \textbf{Traces}
    & \textbf{Events}
    & \textbf{Events}
    & \textbf{MIN}
    & \textbf{AVG}
    & \textbf{MAX} \\\hline
    
L1		&	28,504	&	2.64\%	&	443,862	&	23	&	4	&	15	&	1230\\
L2		&	3,885	&	9.11\%	&	15,096	&	6	&	2	&	3	&	60\\
L3		&	954		&	10.80\%	&	13,740	&	18	&	6	&	14	&	46\\
L4		&	37		&	86.49\%	&	1,156	&	18	&	22	&	31	&	36\\
L5		&	146		&	78.08\%	&	3,764	&	18	&	2	&	25	&	84\\
L6		&	551		&	96.55\%	&	19,174	&	80	&	2	&	34	&	126\\
L7		&	70,512	&	0.28\%	&	830,522	&	8	&	4	&	11	&	40\\
L8		&	9,906	&	2.19\%	&	9,906	&	26	&	6	&	44	&	354\\
L9		&	1,182	&	92.81\%	&	46,282	&	9	&	12	&	39	&	276\\
L10		&	608		&	11.51\%	&	18,238	&	21	&	4	&	2	&	88\\
L11		&	1,214	&	20.18\%	&	11,226	&	12	&	4	&	9	&	58\\
R-Log	&	1,104	&	5.53\%	&	15,468	&	8	&	6	&	14	&	30\\\hline

	\end{tabular}
\egroup
   }}
  	\caption{Descriptive statistics of the logs.}\label{tab:logs}
\end{table}
We implemented SM\textsubscript{2.0} as a Java command-line application,\footnote{Available as ``Split Miner 2.0'' at \url{http://apromore.org/platform/tools}} and we ran the experiments on an Intel Core i7-8565U@1.80GHz with 32GB RAM running Windows 10 Pro (64-bit) and JVM 8 with 14GB of allocated RAM (10GB Stack and 4GB Heap). All the discovery algorithms (SM, SM\textsubscript{2.0}, and IM-lc) were executed using their default input parameters, and we set a timeout of 30 minutes for each algorithm execution and for each measurement.

\vspace{-0.5\baselineskip}
\subsection{Results}

Table~\ref{tab:results} reports the fitness, precision, and simplicity measurements.
Due to space limits, the table does not show the measurements for IMfa
because they were either equal or worse than those for IM-lc, with the exception of those obtained on L9 (which reported a slight improvement).

We can observe that SM\textsubscript{2.0} is less prone to discovering unsound models than SM, with the latter discovering an unsound model every three and the former only discovering sound models. This achievement reflects the effectiveness of our heuristics for removing improper completion.

In terms of accuracy, the results obtained with the alignment-based accuracy and the Markovian accuracy 
are partially consistent in line with previous findings~\cite{augusto2019measuringapproach}.
In fact, the two measuring methods agree on the best models in terms of fitness, precision, and F-score, respectively 100\%, 66\%, and 75\% of the times.

As for fitness, IM-lc outperforms both SM and SM\textsubscript{2.0} as expected~\cite{augusto2018automated}.
In terms of precision and F-score SM\textsubscript{2.0} and SM achieve the highest scores, with SM\textsubscript{2.0} performing better than SM,
most of the times discovering more precise and fitting process models ultimately achieving a higher F-score.
In fact, SM\textsubscript{2.0} accuracy scores are either higher than or equal to those of SM,
the latter outperforming the former in fitness or precision only two times according to the alignment-based accuracy, and only three times according to the Markovian accuracy. Compared to IM-lc, SM\textsubscript{2.0} discovers eleven times more precise process models, indipendently of the measurement method.

As for simplicity, SM\textsubscript{2.0} stands out by producing smaller models than those discovered by both SM and IM-lc (9 times out of 12) and with a lower CFC (10 times out of 12).
However, SM\textsubscript{2.0} and SM cannot systematically produce fully-structured process models as opposed to IM-lc which achieves this by design. Lastly, the execution times of IM-lc, SM, and SM\textsubscript{2.0} are negligible: all the process models were discovered within a minute (except for log L7, where IM-lc timed out).

\begin{figure}[h]
\vspace{-1.0\baselineskip}
	\centering
	\subfloat[SM model discovered from L6.]{
		\includegraphics[width=0.95\textwidth]{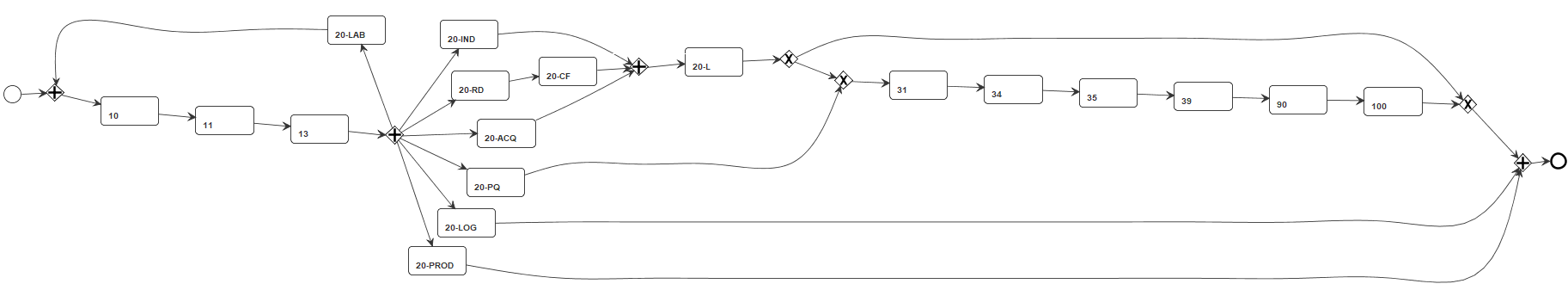}\label{fig:orsm}
	}
	\\
	\subfloat[SM\textsubscript{2.0} model discovered from L6.]{
		\includegraphics[width=0.95\textwidth]{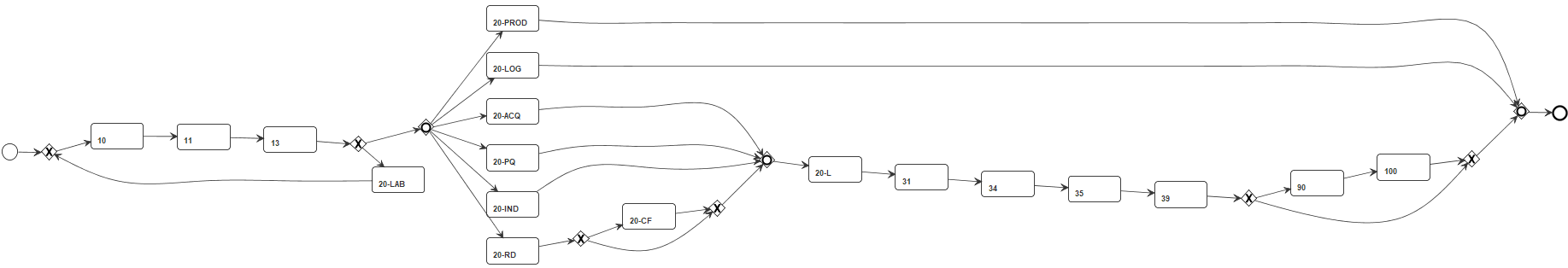}\label{fig:orSMTC}
	}
	\\
	\subfloat[SM model discovered from R-Log.]{
		\includegraphics[width=0.95\textwidth]{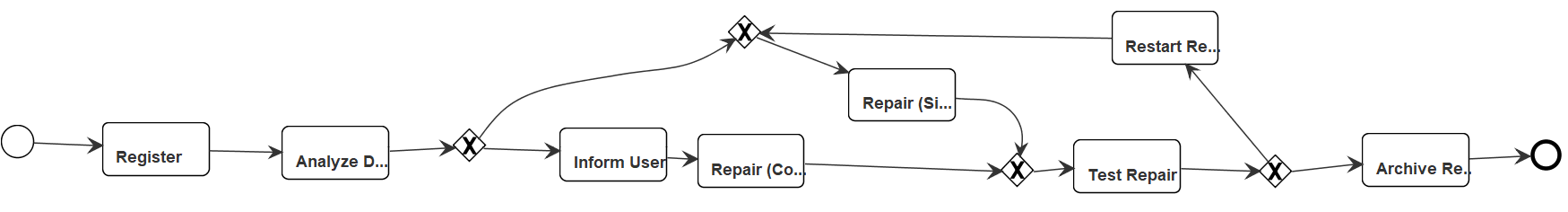}\label{fig:repsm}
	}
	\\
	\subfloat[SM\textsubscript{2.0} model discovered from R-Log.]{
		\includegraphics[width=0.95\textwidth]{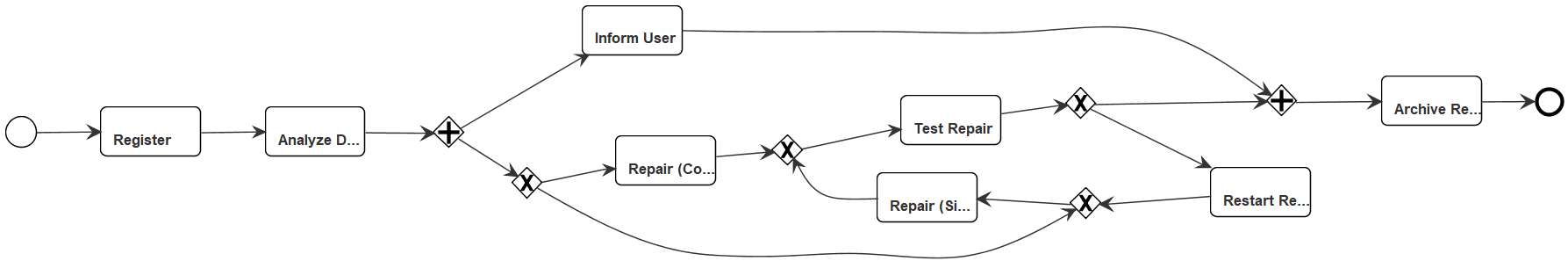}\label{fig:repSMTC}
	}
\caption{Models discovered by SM and SM\textsubscript{2.0} from the L6 and R-Log.}
\label{fig:models}
\end{figure}

\begin{table}
\centering
\bgroup
\def\arraystretch{1.2}
\setlength{\tabcolsep}{2pt}
\begin{tabular}{c|c|ccc|ccc|ccc}

	\textbf{Event}
	& \textbf{Discovery}
	& \multicolumn{3}{c|}{\textbf{Alignment Accuracy}~\cite{adriansyah2012alignment,adriansyah2015measuring}}	
	& \multicolumn{3}{c|}{\textbf{Markovian Accuracy}~\cite{augusto2019measuringapproach}}	
	& \multicolumn{3}{c}{\textbf{Simplicity}}	
    \\\cline{3-11}
    
	\textbf{Log}
	& \textbf{Approach}
	& \textbf{Fitness}
	& \textbf{Precision} 
	& \textbf{F-score}
	& \textbf{Fitness}
	& \textbf{Precision} 
	& \textbf{F-score}
	& \textbf{Size}
	& \textbf{CFC}
	& \textbf{Struct.}
    \\\hline

	&	IM-lc	&	0.88	&	0.78	&	0.83									&	0.82	&	0.15	&	0.25	&	\textbf{40}	&	26	&	\textbf{1.00}\\
L1 	&	SM	&	\textbf{0.98}	&	0.94	&	\textbf{0.96}					&	\textbf{0.96}	&	\textbf{0.44}	&	\textbf{0.60}	&	47	&	32	&	0.47\\
	&	SM\textsubscript{2.0}	&	0.83	&	\textbf{0.97}	&	0.90						&	0.44	&	0.34	&	0.38	&	45	&	\textbf{25}	&	0.56\\\hline

	&	IM-lc	&	0.87	&	0.44	&	0.59									&	0.53	&	0.14	&	0.22 	&	20	&	11	&	\textbf{1.00}\\
L2	&	SM	&	\textbf{0.92}	&	\textbf{1.00}	&	\textbf{0.96}			&	\textbf{0.69}	&	\textbf{0.88}	&	\textbf{0.77}	&	\textbf{14}	&	\textbf{6} &	\textbf{1.00}\\
	&	SM\textsubscript{2.0}	&	\textbf{0.92}	&	\textbf{1.00}	&	\textbf{0.96}		&	\textbf{0.69}	&	\textbf{0.88}	&	\textbf{0.77}	&	\textbf{14}	&	\textbf{6}	&	\textbf{1.00}\\\hline

	&	IM-lc	&	\textbf{0.98}	&	0.71	&	0.82							&	\textbf{0.88}	&	0.08	&	0.14	&	49	&	27	&	\textbf{1.00}\\
L3	&	SM	&	0.96	&	0.97	&	\textbf{0.96}							&	0.72	&	\textbf{0.41}	&	\textbf{0.52}	&	36	&	16	&	0.58\\
	&	SM\textsubscript{2.0}	&	0.93	&	\textbf{0.99}	&	\textbf{0.96}				&	0.40	&	0.07	&	0.12	&	\textbf{31}	&	\textbf{10}	&	0.77\\\hline

	&	IM-lc	&	\textbf{0.98}	&	0.41	&	0.57							&	\textbf{1.00}	&	0.06	&	0.12	&	35	&	12	&	\textbf{1.00}\\
L4	&	SM	&	0.84	&	\textbf{1.00}	&	\textbf{0.91}					&	0.45	&	\textbf{0.79}	&	\textbf{0.57}	&	26	&	6	&	0.46\\
	&	SM\textsubscript{2.0}	&	0.94	&	0.66	&	0.78								&	0.93	&	0.08	&	0.15	&	\textbf{25}	&	\textbf{3}	&	\textbf{1.00}\\\hline

	&	IM-lc	&	\textbf{0.83}	&	0.53	&	0.65							&	\textbf{0.90}	&	0.17	&	0.29	&	33	&	12	&	\textbf{1.00}\\
L5	&	SM	&	\multicolumn{3}{c|}{\textit{unsound}}							&	\multicolumn{3}{c|}{\textit{unsound}}												&	31	&	11	&	0.45\\
	&	SM\textsubscript{2.0}	&	0.76	&	\textbf{0.79}	&	\textbf{0.78}				&	0.86	&	\textbf{0.19}	&	\textbf{0.31}	&	\textbf{27}	&	\textbf{3}	&	0.59\\\hline

	&	IM-lc	&	\multicolumn{3}{c|}{\textit{measurements timeout}}							&	0.10	&	0.00	&	0.01	&	\textbf{126}	&	\textbf{78}	&	\textbf{1.00}\\
L6	&	SM	&	\multicolumn{3}{c|}{\textit{unsound}}		&  \multicolumn{3}{c|}{\textit{unsound}}   &  	161		& 98	&	0.14 \\			
	&	SM\textsubscript{2.0}	&	\textbf{0.70}	&	\textbf{0.66}	&	\textbf{0.68}		&	\textbf{0.31}	&	\textbf{0.23}	&	\textbf{0.26}	&	138		& 80	&	0.50\\\hline

	&	IM-lc	&	\multicolumn{3}{c|}{\textit{discovery timeout}}	&  \multicolumn{3}{c|}{\textit{discovery timeout}}   &  \multicolumn{3}{c}{\textit{discovery timeout}}   \\			
L7	&	SM	&	\textbf{0.88}	&	\textbf{1.00}	&	\textbf{0.94}			&	\textbf{0.73}	&	\textbf{0.90}	&	\textbf{0.81}	&	\textbf{12}	&	\textbf{2}	&	\textbf{1.00}\\
	&	SM\textsubscript{2.0}	&	\textbf{0.88}	&	\textbf{1.00}	&	\textbf{0.94}		&	\textbf{0.73}	&	\textbf{0.90}	&	\textbf{0.81}	&	\textbf{12}	&	\textbf{2}	&	\textbf{1.00}\\\hline

	&	IM-lc	&	\textbf{0.85}	&	0.40	&	0.55							&	\textbf{0.87}	&	0.03	&	0.06	&	61	&	39	&	\textbf{1.00}\\
L8	&	SM	&	\multicolumn{3}{c|}{\textit{unsound}}							& \multicolumn{3}{c|}{\textit{unsound}}										&	160	&	118	&	0.02\\
	&	SM\textsubscript{2.0}	&	0.77	&	\textbf{0.76}	&	\textbf{0.77}				&	0.38	&	\textbf{0.33}	&	\textbf{0.35}	&	\textbf{46}	&	\textbf{26}	&	0.70\\\hline

	&	IM-lc	&	\textbf{0.94}	&	0.26	&	0.41							&	\textbf{0.92}	&	0.43	&	\textbf{0.58}	&	23	&	11	&	\textbf{1.00}\\
L9	&	SM	&	\multicolumn{3}{c|}{\textit{unsound}}							&	 \multicolumn{3}{c|}{\textit{unsound}}									&	\textbf{17}	&	\textbf{5}	&	0.53\\
	&	SM\textsubscript{2.0}	&	0.57	&	\textbf{0.91}	&	\textbf{0.70}				&	0.28	&	\textbf{0.45}	&	0.35	&	\textbf{17}	&	\textbf{5}	&	0.47\\\hline

	&	IM-lc	&	\textbf{0.95}	&	0.75	&	0.84							&	\textbf{0.98}	&	0.15	&	0.26	&	31	&	8	&	\textbf{1.00}\\
L10	&	SM	&	0.77	&	\textbf{1.00}	&	\textbf{0.87}					&	0.95	&	\textbf{0.93}	&	\textbf{0.94}	&	\textbf{29}	&	\textbf{6}	&	\textbf{1.00}\\
	&	SM\textsubscript{2.0}	&	0.77	&	\textbf{1.00}	&	\textbf{0.87}				&	0.95	&	\textbf{0.93}	&	\textbf{0.94}	&	\textbf{29}	&	\textbf{6}	&	\textbf{1.00}\\\hline

	&	IM-lc	&	\textbf{0.91}	&	0.75	&	0.82							&	\textbf{0.45}	&	0.14	&	0.22	&	36	&	21	&	\textbf{1.00}\\
L11	&	SM	&	0.83	&	\textbf{0.90}	&	\textbf{0.87}					&	0.29	&	0.26	&	\textbf{0.28}	&	44	&	28	&	0.16\\
	&	SM\textsubscript{2.0}	&	0.84	&	\textbf{0.90}	&	\textbf{0.87}				&	0.06	&	\textbf{0.33}	&	0.10	&	\textbf{22}	&	\textbf{11}	&	0.59\\\hline

		&	IM-lc	&	\textbf{0.99}	&	0.98	&	\textbf{0.99}				&	\textbf{1.00}	&	0.96	&	\textbf{0.98}	&	16	&	5	&	\textbf{1.00}\\
R-Log	&	SM	&	0.91	&	\textbf{0.99}	&	0.95						&	0.45	&	0.83	&	0.59	&	\textbf{14}	&	\textbf{4}	&	0.36\\
		&	SM\textsubscript{2.0}	&	0.98	&	0.97	&	0.98							&	0.94	&	\textbf{0.98}	&	0.96	&	16	&	5	&	0.50\\\hline
	
	\end{tabular}
\egroup
  	\caption{Experiment results.}\label{tab:results}
\end{table}
Figure~\ref{fig:models} shows two qualitative examples of the improvements yielded by SM\textsubscript{2.0}.
Considering the models from the L6 log (Figures~\ref{fig:orsm} and~\ref{fig:orSMTC}), SM\textsubscript{2.0} discovered the inclusive-OR relations between several activities of the process and removed the improper completion, while SM produced an unsound model. In the specific case of the L6 log, we also had the chance to validate the discovered model with the process analysts of the organization this log was extracted from, who confirmed that the activities were indeed in an inclusive-OR relation. Considering the models from the R-log (Figures~\ref{fig:repsm} and~\ref{fig:repSMTC}), 
only SM\textsubscript{2.0} discovers the concurrency relations between its activities, while SM mixes us the concurrency relations with loops.

\vspace{-1.0\baselineskip}


\vspace{-1.5\baselineskip}
\section{Conclusion}\label{sec:conclusion}

In this paper, we presented Split Miner 2.0 (SM\textsubscript{2.0}), an extension of Split Miner (SM)~\cite{augusto2018split} that exploits the activities' start and end timestamps recorded in an event log to discover true concurrency and inclusive choice relations between activities.
This is achieved by redesigning the discovery of a directly-follows graph from an event log, adapting the concurrency notion of Van der Werf et al.~\cite{van2013mining}, and introducing an intuitive heuristic to identify inclusive relations.
Furthermore, given that SM cannot guarantee sound process models, we designed an heuristic that reduces the chances of discovering process models exhibiting improper completion.
The empirical evaluation shows that SM\textsubscript{2.0} can discover more concurrent relations than SM, remove improper completion, and identify OR-splits, while preserving SM's model accuracy and reducing the complexity.

Although several studies have investigated the problem of automated process discovery from event logs~\cite{augusto2018automated},
most of them operate on simple event logs with only three attributes: case id, timestamp, and activity label.
Future research work in this area may focus on designing more sophisticated automated process discovery algorithms that can discover more complex BPMN process models by leveraging additional information that may be available in real-life event logs.
Another direction for future work is to design accuracy measures such as fitness and precision that go beyond simple control-flow relations and include support for inclusive gateways, including the OR-join.

\medskip\noindent\textbf{Acknowledgments.} Research funded by the Australian Research Council (grant DP180102839) and the Estonian Research Council (grant PRG887).
\vspace{-0.5\baselineskip}

\bibliographystyle{plain}
\bibliography{lit}

\end{document}